# Physics of collective cell migration

Ivana Pajic-Lijakovic* and Milan Milivojevic

Faculty of Technology and Metallurgy, Belgrade University, Karnegijeva 4, Belgrade, Serbia

Correspondence to Ivana Pajic-Lijakovic (iva@tmf.bg.ac.rs)

**Abstract**

Movement of cell clusters along extracellular matrices (ECM) during tissue development, wound healing, and early stage of cancer invasion involve various inter-connected migration modes such as: (1) cell movement within clusters, (2) cluster extension (wetting) and compression (de-wetting), and (3) directional cluster movement. It has become increasingly evident that dilational and volumetric viscoelasticity of cell clusters and their surrounding substrate significantly influence these migration modes through physical parameters such as: cell and matrix surface tensions, interfacial tension between cells and substrate, gradients of surface and interfacial tensions, as well as, the accumulation of cell and matrix residual stresses. Inhomogeneous distribution of cell surface tension along migrating cell cluster can appear as a consequence of different strength of cell-cell adhesion contacts and cell contractility between leader and follower cells. This inhomogeneity in the form of cell surface tension gradient influences cell movement within the cluster. However, this phenomenon has not been considered yet. The cell residual stress accumulation can reduce movement of epithelial-like cells, while physical properties of ECM such as the matrix surface tension gradient and matrix residual stress accumulation are responsible for directional cell migration. An inhomogeneous accumulation of the matrix stress leads to the establishment of a matrix stiffness gradient which guides directional cell migration. While the directional cell migration caused by the matrix stiffness gradient (i.e. durotaxis) has been widely elaborated, the structural changes of matrix surface caused by cell tractions which lead to the generation of the matrix surface tension gradient has not been considered yet.

The main goal of this theoretical consideration is to clarify the roles of various physical parameters in collective cell migration based on the formulating biophysical model. This complex phenomenon is discussed on the model systems such as the movement of cell clusters on the collagen I gel matrix by simultaneously reviewing various experimental data with and without cells.

1. Introduction

Collective cell migration is an essential process during morphogenesis, wound healing, and cancer invasion (Clark et al., 2015;2022; Barriga and Mayor, 2019). Movement of cell collectives frequently occurs in a highly directional manner (Shellard and Mayor, 2020). Directional cell movement, i.e. taxis is induced by various chemical, mechanical, and electrical stimuli. Consequently, established gradient of: (1) nutrient concentration induces chemotaxis, (2) electric field induces galvanotaxis, (3) matrix stiffness induces durotaxis, and (4) cellular adhesion sites or substrate-bound cytokines induces haptotaxis (Murray et al., 1988; Shellard and Mayor, 2020; 2021). Directional cell migration has been mainly considered in 2D by monitoring free expansion of cell monolayers and movement of cell clusters on substrate matrix (Serra-Picamal et al., 2012; Nnetu et al., 2012;2013; Clark et al., 2022). Among others, collagen I gel has been widely used as a substrate matrix. It is in accordance with the fact that this type of network represents a constituent of stroma. During early stage of epithelial cancers, cell clusters migrate along the stroma which is composed primarily of collagen I extracellular matrix (ECM) (Clark et al., 2015).

Cells generate mechanical forces on ECM in the range of ~10-100 nN during their movement which occurs at long time scale (i.e. a time scale of hours) (Hall et al., 2016; Steinwachs et al., 2016; Emon et al., 2021). These forces are much larger than the necessary force for breaking electrostatic and hydrophobic bonds in collagen I networks, which is equal to ~20 $pN$ (Nam et al., 2016). The force of a few $nN$ is enough for stretching of the collagen filament up to 20% strain (Gautieri et al., 2012). This means that cells are capable of inducing significant volumetric and surface structural rearrangement of collagen I gel which feeds back on the migration persistence (Clark et al., 2022). Altered volumetric rearrangement of the collagen I matrix results in the matrix stiffening, while the surface rearrangement influences the matrix surface tension. The volumetric and surface rearrangement of collagen I gel are inter-related based on the Young-Laplace equation (Pajic-Lijakovic and Milivojevic 2023). Cell clusters can migrate persistently on collagen I matrix, governed by physical mechanisms, without the establishment of the front-rear polarization (Clark et al., 2022). In this case, the migration persistence is caused by structural changes of matrix related to dilational and volumetric viscoelasticity (Pajic-Lijakovic et al., 2022d). The dilational viscoelasticity describes the change of surface energy of collagen I gel in the form of the matrix dynamic surface tension caused by cell tractions. Volumetric structural changes of matrix induce the matrix residual stress accumulation and on that base results in the matrix stiffening (Pajic-Lijakovic et al., 2022d). While the directional cell movement caused by the matrix stiffness gradient (i.e. durotaxis) has been intensively studied (Sunyer et al., 2016), the influence of matrix surface tension change on the directional movement of cell clusters has been less elaborated. The main goal of this review is to discuss the influence of the physical parameters such as: matrix surface tension, cell surface tension, cell-matrix interfacial tension accompanied by their gradients, viscoelasticity of the cell cluster and viscoelasticity of the collagen I matrix on: (1) cell rearrangement within a cell cluster, (2) cell cluster extension (wetting) or compression (de-wetting), and (3) directional movement of cells. These modes of cell migration are inter-connected (Pallarès et al., 2022). Cell clusters de-wet soft substrate and wet stiff one during their movement (Pallarès et al., 2022). The aim of this consideration is to clarify the inter-connection among the migration modes by discussing the role of introduced physical parameters.

Movement of cell clusters on collagen networks causes complex in-plane and out-of-plane strains which feeds back the movement itself on a complex way. Clark et al. (2022) revealed that cell clusters exert asymmetric inward-facing radial traction forces near the cluster edge which induce an in-plane extension of the collagen network in regions surrounding the cells and network in-plane compression in the region directly under the cell cluster. Besides in-plane strain, out-of-plane strain is generated caused by downward-facing tractions in the middle of the cluster (Pajic-Lijakovic et al., 2022d). Induced in-plane strain results in surface structural changes of the matrix, while out-of-plane strain results in volumetric structural changes of the collagen I matrix which leads to the matrix stiffening. The surface structural changes of matrix include extension of collagen fibers and their radial alignment around the cell cluster, which lead to change in the collagen concentration under the cluster and around the cluster (Clark et al., 2022). The resulted distribution of collagen concentration around the cell cluster is asymmetric such that the collagen concentration near the cluster front region is ~30% times lower than the one near the cluster rear (Clark et al., 2022). Change of the collagen concentration is caused by the in-plane strain which results in an establishment of the matrix surface tension gradient. However, the impact of this gradient on the directional cell migration has not been elaborated yet. The dependence of collagen I surface tension on the collagen concentration has been considered on collagen I films in experiments without cells (Kezwon and Wojciechowski, 2014). It would be interesting to compare dilational viscoelasticity of collagen I film and fibrinogen film under the same conditions in experiments without cells. Both of them are widely used as substrate matrices for cell migration assays. An interesting result can be extracted. While the protein concentration increase in the range of $1 \frac{mg}{ml} - 4 \frac{mg}{ml}$ induces a decrease in the surface tension of collagen networks, it has no effect on the surface tension fibrinogen networks (Gudapati et al., 2020). Consequently, it seems that the surface structural changes of the matrix, caused by cell movement, can have an impact on directional cell migration on collagen I matrix, but have no effect on cell movement on the fibrinogen matrix.

The focus of this review is to consider: (1) the cell rearrangement within the cluster, (2) cell extension (wetting)/ compression (de-wetting) during the cluster movement, and (3) directional cell movement based on physical parameters such as: cell and matrix surface tensions, interfacial tension between cells and matrix, as well as, the corresponding gradients of surface and interfacial tensions, and the viscoelasticity of cell cluster in one hand and the viscoelasticity of the matrix on the other. In addition, we point out the importance of accounting for the surface characteristics of tissue and surrounding ECM in advancing cancer physics research, and discuss open problems and potential opportunities that can be addressed with these tools. We also present a new biophysical model in order to point out the role of these physical parameters in rearrangement and movement of cell collectives by considering the model system such as movement of cell clusters on collagen I network.

## 2. Physical parameters which influence movement of cell cluster on collagen I matrix

Various modes of cell cluster movement such as: (1) the cell movement within the cluster, (2) cluster wetting/de-wetting, and (3) directional cell movement are shown schematically in **Figure 1**.

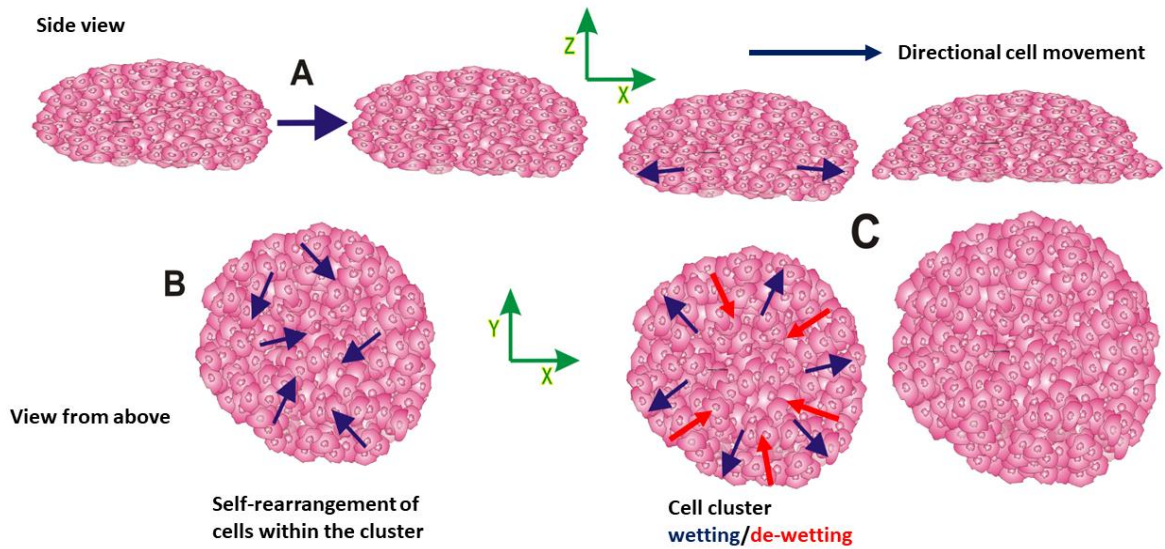

**Figure 1.** Schematic presentation of migration modes such as: (1) the cell movement within the cluster, (2) cluster wetting/de-wetting, and (3) directional cell movement.

These migration modes depend on interplay between physical parameters such as: (1) cell surface tension, (2) matrix surface tension, (3) cell-matrix interfacial tension, (4) gradients of surface and interfacial tensions, (5) cell residual stress, and (6) matrix residual stress influences the rearrangement and movement of cell cluster on collagen I matrix. The main characteristics of these parameters are inhomogeneous distributions near the cell-matrix biointerface and time-dependence. Some physical parameters influence the movement of cells: (1) directly and (2) indirectly by influencing the cell and matrix residual stresses which have a feedback on cell packing density and velocity (Pajic-Lijakovic et al., 2022e). The inter-relation among physical parameters which guide collective cell migration is shown in **Figure 2**.

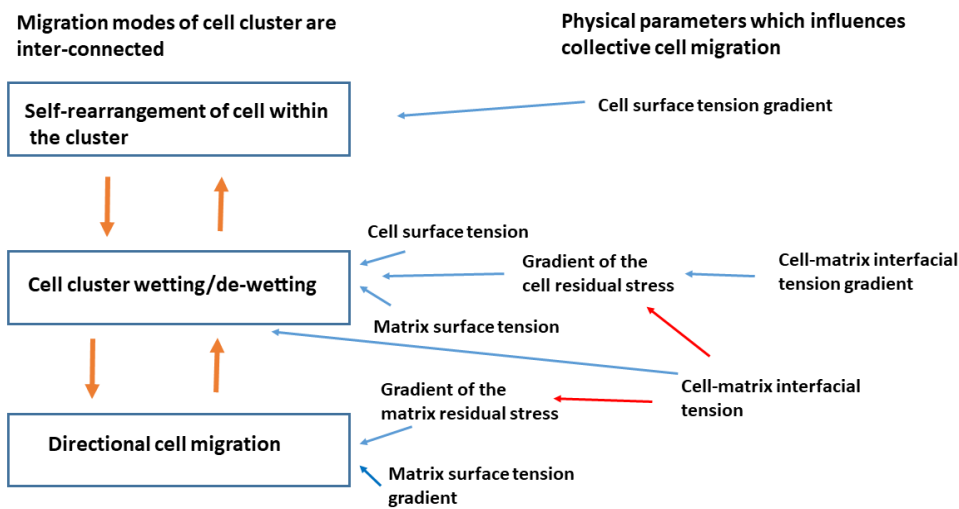

**Figure 2**. The inter-relation among physical parameters which guide collective cell migration.

In order to discuss this complex phenomenon in the context of the formulated biophysical model, it is necessary to describe these physical parameters in more details.

**2.1 Dilational and volumetric viscoelasticity of collagen I matrix: the directional cell movement**

The in-plane and out-of-plane strains, caused by movement of cell cluster, lead to an establishment of the matrix surface tension gradient and matrix stiffness gradient, respectively, which are responsible for the directional cell movement as was shown schematically in **Figure 3**.

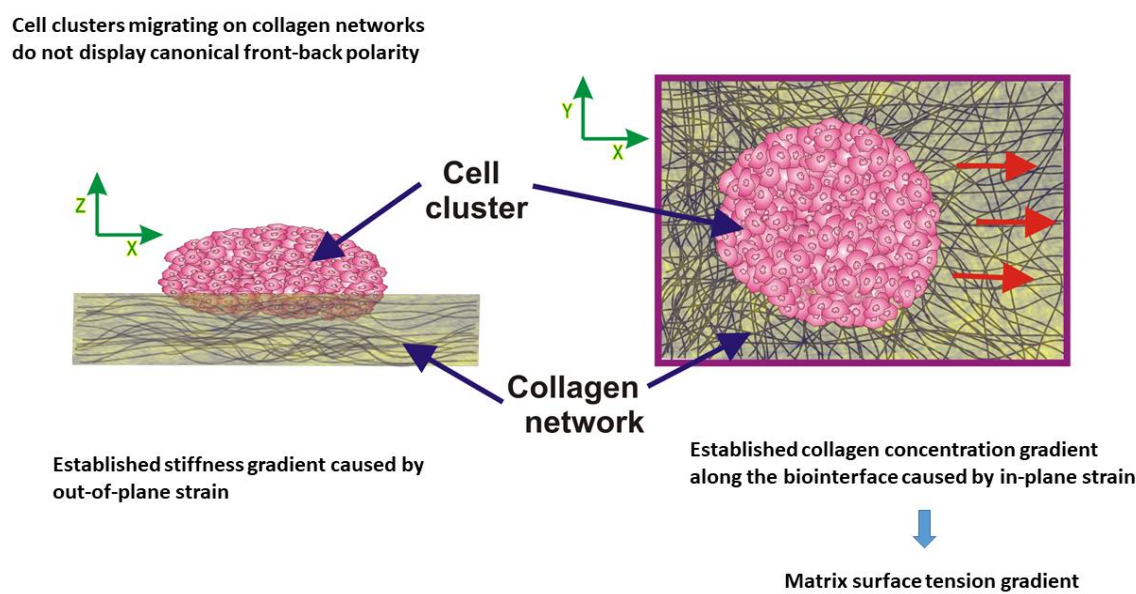

**Figure 3.** Schematic presentation of the rearrangement of collagen I matrix caused by movement of cell cluster.

The altered structural changes of collagen I networks occur at various time scales from milliseconds to hours (Pryse et al., 2003; Gautieri et al., 2012; Nam et al., 2016; Pajic-Lijakovic et al. 2022d). In order to understand better the relationships between (1) in-plane strain and the matrix surface tension change, (2) out-of-plane strain and the matrix residual stress accumulation which results in the matrix stiffening, and (3) the matrix surface tension and matrix normal residual stress accumulation, it is necessary to point to the main characteristics of the dilational and volumetric viscoelasticity of collagen I networks obtained in various experiments without cells.

The surface tension of collagen I film relaxes on change in the surface area $\Delta A_M$ which has been considered by Kezwon and Wojciechowski (2014) in experiments without cells. The relaxation time corresponds to a few minutes, while the necessary time for reaching out the surface tension equilibrium state $\gamma_{Meq}$ is more than 1 h at 21 $^oC$ (Kezwon and Wojciechowski, 2014). The relaxation time increases with the collagen concentration, while the collagen surface tension decreases with the collagen concentration. An increase in the collagen concentration from $1\ \frac{mg}{ml}$ to $4\ \frac{mg}{ml}$ (in the experiments without cells) causes a decrease in the equilibrium collagen I surface tension $\gamma_{Meq}$ from $62\ \frac{mN}{m}$ to $56\ \frac{mN}{m}$ at 21 $^oC$ (Kezwon and Wojciechowski, 2014). It is not clear whether this surface tension difference is large or small. To clarify this issue, we would like to discuss the corresponding surface tension difference of 6 $\frac{mN}{m}$ for some well-known system such as water. An increase in the temperature from 0 $^oC$ to 50 $^oC$ induces a decrease in the surface tension of water for 7 $\frac{mN}{m}$

An indicated range of the collagen concentration also corresponds to various experiments which have been performed with cells. While the equilibrium collagen surface tension decreases with the concentration, the collagen surface tension change $\Delta\gamma_M$ during the relaxation process, caused by changing the surface area $\frac{\Delta A_M}{A_M}$, increases with the concentration (where $\Delta\gamma_M = \gamma_{M0} - \gamma_{Meq}$ and $\gamma_{Meq}$ is the equilibrium matrix surface tension). The phenomenon is pronounced at higher temperature of 34.5 $^oC$ (Kezwon and Wojciechowski, 2014). Consequently, the equilibrium surface tension of collagen I film satisfies the condition $\gamma_{Meq} \sim C_{col}^{-1}$ (where $C_{col}$ is the collagen surface concentration). It is in accordance with fact that collagen fibers are capable of establishment inter-fiber connections in the form of physical gel-like structures at higher collagen concentrations and higher temperatures (Nam et al., 2016). The relationship between the collagen surface tension change $\Delta\gamma_M$ and surface area $\frac{\Delta A_M}{A_M}$ can be formulated based on some suitable constitutive model of the dilational viscoelasticity $\Delta\gamma_M = \Delta\gamma_M(\frac{\Delta A_M}{A_M})$ (Babak et al., 2005). The Fourier transform of the constitutive model can be presented in the form of $F(\Delta\gamma_M) = E_{sM}^* F\left(\frac{\Delta A_M}{A_M}\right)$ (where $F(\cdot)$ is the Fourier operator, $E_{sM}^*(\omega)$ is the dynamic surface modulus of the matrix equal to $E_{sM}^*(\omega) = E_{sM}'(\omega) + iE_{sM}''(\omega)$, while $E_{sM}'$ is the surface storage modulus, $E_{sM}''$ is the surface loss moduli, $\omega$ is the angular velocity, and $i = \sqrt{-1}$). The surface storage modulus represents a measure of the storage energy within the surface area, while the surface loss modulus represents a measure of the energy dissipation caused by the structural changes of the surface area $\frac{\Delta A_M}{A_M}$. The main characteristic of the dilational viscoelasticity of collagen I surfaces without cells, for the concentration range 1 $\frac{mg}{ml}$ – 4 $\frac{mg}{ml}$, is that $E_{sM}' > E_{sM}''$ which points to the viscoelastic solid behaviour (Kezwon and Wojciechowski, 2014).

The surface tension of collagen I sample influences the residual stress accumulation under external strain conditions. The corresponding matrix normal residual stress can be expressed based on the Young-Laplace equation as: $\widetilde{\boldsymbol{\sigma}}_{MrV} = \Delta p_M \widetilde{\boldsymbol{I}} + \widetilde{\boldsymbol{\sigma}}_{MrV}^d$ (where $\Delta p_M \widetilde{\boldsymbol{I}}$ is the isotropic part of stress equal to $\Delta p_M = -\gamma_M(\vec{\boldsymbol{\nabla}} \cdot \vec{\boldsymbol{n}})$, $\vec{\boldsymbol{n}}$ is the normal vector on the surface, $\widetilde{\boldsymbol{I}}$ is the unit tensor, and $\widetilde{\boldsymbol{\sigma}}_{MrV}^d$ is the deviatoric part of stress caused by external strain). The residual stress within collagen I network increases during successive stress relaxation cycles under constant uni-axial extensional strain per cycle from $\sim 5\ Pa$ after first relaxation cycle to $\sim 35\ Pa$ after the third relaxation cycle in the experiments without cells (Pryse et al., 2003). In experiments with cells, the residual stress accumulation within the collagen network is responsible for the matrix stiffening and the establishment of the stiffness gradient, which can have a feedback on the cell movement persistence (Pajic-Lijakovic et al. 2022d).

After discussing dilational and volumetric viscoelasticity of collagen I networks obtained in experiments without cells, it is necessary to point to the structural changes of collagen I network matrix caused by movement of cell cluster on the matrix surface. The cell movement induces the matrix in-plane extension around the cluster edge, pronounced around the cluster front, which leads to extension of collagen fibers and their radial alignment around the cluster edge (Clark et al., 2022). Consequently, the gradient of the collagen surface concentration, caused by the in-plane strain leads to an establishment of the matrix surface tension gradient such that $\gamma_M^F > \gamma_M^R$ (where the superscript $R$ is the cluster rear region and the superscript $F$ is the front region). The corresponding matrix surface tension gradient can contribute to the cluster migration persistence. In this context, we can provide two arguments:

- Established gradient of the matrix surface tension $\vec{\nabla}\gamma_M$ drives movement of partially disconnected collagen fibers from the region of lower surface tension (at the cluster rear) to the region of larger surface tension (at the cluster front). This phenomenon represents a part of the Marangoni effect recognizable in various soft matter systems (Karbalaei et al., 2016) in which the surface tension gradient is induced by change in temperature or surface concentration of the system constituents.
- Larger matrix surface tension corresponds to a higher cell spreading coefficient for the same cell surface tension and cell-matrix interfacial tension (Pajic-Lijakovic and Milivojevic 2023). The spreading coefficient is expressed as: $S^{c-M} = \gamma_M - (\gamma_c + \gamma_{Mc})$ (where $\gamma_c$ is the surface tension of cell aggregate and $\gamma_{Mc}$ is the cell-matrix interfacial tension). Detail description of this complex phenomenon will be given in next sections.

Cell cluster also induces out-of-plane compression of collagen I matrix, pronounced under the cluster central part, caused by cell tractions which contribute to the deviatoric part of the normal matrix residual stress accumulation within the matrix while the isotropic part of the matrix normal residual stress is generated by work of the matrix surface tension based on the Young-Laplace equation (**Table 1**) (Pajic-Lijakovic and Milivojevic, 2023). The matrix residual stress distribution can induce the establishment of the matrix stiffness gradient (Pajic-Lijakovic et al., 2022d).

Besides matrix surface tension, the cell surface tension accompanied the cell-matrix interfacial tensions and their gradients govern the movement of cell clusters on collagen I matrix. For deeper understanding the phenomenon of cell migration, it is necessary to discuss these parameters which will be incorporated within the biophysical model.

**2.2 Surface tension of cell clusters: the movement of cells within the cluster**

Macroscopic surface tension of cell clusters is time dependent physical parameter which represents a measure of cluster cohesiveness. This surface tension is influenced by the state of single cells and the extension or compression of multicellular system caused by collective cell migration (Guevorkian et al., 2021; Pajic-Lijakovic and Milivojevic, 2023). The state of single cells includes the cell contractility and strength of cell-cell adhesion contacts. Contractile epithelial cells have larger surface tension than non-contractile ones (Devanny et al., 2021). In this context, two reasons can be provided: (1) contractile cells adsorbed contractile energy and become stiffer than non-contractile ones (Pajic-Lijakovic and Milivojevic, 2022a) and (2) cell contractility enhances the strength of E-cadherin mediated adherens junctions (AJs) (Devanny et al., 2021). The surface tension of epithelial-like systems are lower than the surface tension of collagen I matrix, i.e. $\gamma_c < \gamma_M$ (Pajic-Lijakovic et al., 2022e). The tissue surface tension varies significantly for various cellular systems. The surface tension of: (1) F9 WT cell aggregates is $4.5\ \frac{mN}{m}$ (Stirbat et al., 2013), (2) MCF 10-A aggregates is $45 \pm 18\ \frac{mN}{m}$ (Nagle et al., 2022), and (3) aggregates of CHO cells $22.8 \pm 3\ \frac{mN}{m}$ (Efremov et al., 2021). The surface tension of collagen matrix obtained for the collagen concentration of $4\ \frac{mg}{ml}$ is significantly larger than the tissue surface tension and equal to $56\ \frac{mN}{m}$ (Kezwon and Wojciechowski, 2014).

Extension of epithelial surfaces lead to a significant increase in a cell surface tension (Guevorkian et al., 2021). It is in accordance with fact that the extension enhances the strength of E-cadherin mediated adherens junctions (Devanny et al., 2021). Local extension of the murine sarcoma (S180) aggregate surface by applying the micropipette aspiration force in the range of 0.5 $\mu N$ to 1.5 $\mu N$ leads to an increase in the surface tension from $\gamma_c \sim 7 \frac{mN}{m}$ to $\sim 22 \frac{mN}{m}$ (Guevorkian et al., 2021). Collective cell migration also induces successive extension and compression of multicellular surfaces in the form of mechanical waves (Serra-Picamal et al., 2012; Notbohm et al., 2016; Pajic-Lijakovic and Milivojevic, 2020;2022b). However, multicellular surfaces haven't been considered in the context of dilational viscoelasticity yet. We can provide here only a qualitative analysis based on some experimental findings in the context of the relationship between the cell surface tension change $\Delta \gamma_c$ induced by changing the multicellular surface area $\frac{\Delta A_c}{A_c}$ (where $\Delta \gamma_c = \gamma_c - \gamma_{c\,eq}$, $\gamma_c$ is the tissue surface tension, and $\gamma_{c\,eq}$ is the equilibrium tissue surface tension). This change in the surface tension is caused by (1) change in the number of cells per surface area which has a feedback on the strength of AJs (Pajic-Lijakovic and Milivojevic, 2022a;2023), and (2) change in the surface area per single cells caused by change of the cluster surface area (Guevorkian et al., 2021). It would be interesting to extract the dynamic complex surface modulus of cells $E_{sc}^*(\omega) = \frac{F(\Delta \gamma_c)}{F\left(\frac{\Delta A_c}{A_c}\right)}$ (where $F(\cdot)$ is the Fourier operator, $E_{sc}'(\omega)$ is the cell surface storage modulus, and $E_{sc}''(\omega)$ is the cell surface loss modulus) and consider the ratio between storage and loss moduli, i.e. $\frac{E_{sc}'(\omega)}{E_{sc}''(\omega)}$ under various experimental conditions, in order to extract more information relevant for collective cell migration.

The cell surface tension varies along the cell cluster. Protrusive leading edge would be expected to have higher surface tension than the cluster central part consists of follower cells which are less contractile. It is in accordance with the fact that Rac1 appears to be down regulated at cell-cell junctions at the cluster interior (Hidalgo-Carcedo et al., 2011; Clark et al., 2022). Cell contractility enhances the strength of E-cadherin mediated adherens junctions (Devanny et al., 2021). Consequently, the surface tension of active, contractile epithelial-like cells is larger than the surface tension of non-contractile ones (Devanny et al., 2021). Based on these findings, we can conclude that the cell surface tension established at the cluster front is larger than that at the cluster interior. The established cell surface tension gradient governs cell movement from the region of lower cell surface tension (characteristic for the follower cells) to the region of larger cell surface tension (characteristic for leader cells) (Pajic-Lijakovic and Milivojevic, 2022c). This phenomenon represents also a part of the Marangoni effect recognizable in various soft matter systems which directs the movement of the system constituents from the region of lower surface tension to the region of larger surface tension (Karbalaei et al., 2016). This effect is also responsible for cell segregation within co-cultured cell clusters (Maître et al., 2012). Consequently, this physical mechanism accompanied by biochemical mechanism related to cell signaling (Clark et al., 2022) influence cell self-rearrangement within the cluster.

**2.3 Cell-matrix interfacial tension**

The cell-matrix interfacial tension and its gradient influence the modes of cell migration directly through the spreading factor and indirectly by influencing the accumulation of cell and matrix residual stresses as was shown schematically in **Figure 2**.

The cell-matrix interfacial tension has not been measured yet. The interfacial tension $\gamma_{cM}$ can be expressed as:

$$\gamma_{cM} = \gamma_c + \gamma_M - \omega_a \tag{1}$$

where $\omega_a$ is the adhesion energy between cells and matrix which can be expressed as: $\omega_a = \frac{1}{A_{int}} \sum_{i=1}^{N} \frac{1}{2} k_c |\vec{u}_M^{\ c}|^2_i$ (Murray et al., 1988), $N$ is the number of focal adhesions (FAs) within the interfacial area $A_{int}$, $k_c$ is the elastic constant per single FA, and $\vec{u}_M^{\ c}$ is the matrix displacement field caused by cell tractions. Clark et al. (2022) revealed that cell clusters preform asymmetric tractions during their movement on collagen I matrix such that the maximum tractions is induced at the cluster rear which pointed out that $\omega_a^F < \omega_a^R$. Accordingly with previously extracted conclusion that: (1) $\gamma_c^F > \gamma_c^R$ caused by inhomogeneous cell extension and contractility within the cluster, (2) $\gamma_M^F > \gamma_M^R$ caused by change in the collagen concentration, and (3) $\omega_a^F < \omega_a^R$ caused by asymmetric cell tractions, and based on eq. 1, we can conclude that the interfacial tension is larger at the cluster front and decrease toward the cluster rear, i.e. $\gamma_{cM}^F > \gamma_{cM}^R$. Besides, cell matrix interfacial tension, the cell and matrix surface tensions also contribute to the cluster extension (wetting) or compression (de-wetting) expressed in the form of the cell spreading coefficient.

## 2.4 Physical parameters responsible for cell cluster wetting/de-wetting

Multicellular systems perform oscillatory extension (wetting) and compression (der-wetting) during collective cell migration which have been discussed in the context of mechanical waves (Serra-Picamal et al., 2012; Notbohm et al., 2016; Pajic-Lijakovic and Milivojevic, 2020). The phenomenon has been recognized in various 2D and 3D model systems such as: (1) the free expansion of cell monolayers (Serra-Picamal et al., 2012), (2) rearrangement of confluent cell monolayers (Notbohm et al., 2016), (3) cell aggregate rounding after uni-axial compression (Mombash et al., 2005; Pajic-Lijakovic and Milivojevic, 2022b), (4) cell aggregate wetting/de-wetting on rigid substrate (Beauene et al., 2018), and (5) fusion of two cell aggregate (Pajic-Lijakovic and Milivojevic, 2023a).

Oscillatory wetting and de-wetting of cell clusters are guided by surface and volumetric physical parameters. Surface parameter in the form of the cell spreading coefficient represents interplay between cell and matrix surface tensions accompanied by the interfacial tension between them. Volumetric parameters are the cell and matrix residual stresses (normal and shear) which represents a consequence of cell-matrix interactions at the biointerface. Consequently, an interfacial tension influences both surface and volumetric physical parameters. We will discuss this relationship in more details within next two sections.

### 2.4.1 Cell spreading coefficient

The cell-matrix interfacial tension accompanied by the cell and matrix surface tensions contributes to the extension (wetting) or compression (de-wetting) of the cell cluster during its movement. The corresponding spreading coefficient can be expressed as: $S^{c-M} = \gamma_M - (\gamma_c + \gamma_{Mc})$ (Pajic-Lijakovic and Milivojevic, 2023). Two cases can be distinguished in the context of the spreading coefficient such as:

(1) $S^{c-M} > 0$ which means that $\gamma_M > \gamma_c$ and $\gamma_M > \gamma_{cM}$ and corresponds to the cell cluster extension and

(2) $S^{c-M} < 0$ which means that $\gamma_M < \gamma_c + \gamma_{cM}$ and corresponds to the cell cluster compression (Pajic-Lijakovic and Milivojevic, 20223).

While the cluster front performs oscillatory extension during its movement on collagen matrix, the cluster rear could undergo extension or compression depending on the cell type and rheological behaviour of the substrate matrix (Beaune et al., 2018; Alert and Casademunt, 2018). The cell-matrix interfacial tension influences the residual stress accumulation within migrating cell cluster and within matrix which will be discussed based on the modelling consideration.

**2.4.2 The residual stress accumulation within cell cluster and within collagen I matrix**

Residual stresses are defined as self-equilibrating stresses which exist in materials even in the absence of external loads. In the viscoelastic materials, this stress can be dissipative or elastic depending primarily on the strength of cell-cell adhesion contacts. The corresponding shear and normal residual stresses for migrating cell cluster and collagen I matrix are shown in **Table 1**.

**Table 1**. The residual stress accumulation within cell cluster and substrate matrix

| Residual stress | Cell cluster | Collagen I matrix |
|---|---|---|
| Normal residual stress | $\tilde{\sigma}_{crV} = \pm \Delta p_{c \to M} \tilde{I} + \tilde{\sigma}_{crV}{}^d$<br><br>Isotropic parts of the normal stress:<br>$\Delta p_{c \to M} = -\gamma_{cM} (\vec{\nabla} \cdot \vec{n})$<br>Deviatoric part of the cell normal stress $\tilde{\sigma}_{crV}{}^d$:<br>$\tilde{\sigma}_{crV}{}^d = \tilde{\sigma}_{crV}{}^{CCM} + \tilde{\sigma}_{cV}{}^G$ | $\tilde{\sigma}_{MrV} = \mp \Delta p_{c \to M} \tilde{I} + \tilde{\sigma}_{MrV}{}^d$<br><br>Deviatoric part of the matrix normal stress:<br>$\tilde{\sigma}_{MrV}{}^d = \tilde{\sigma}_{MrV}{}^{TR} + \tilde{\sigma}_{MV}{}^G$ |
| Shear residual stress | $\tilde{\sigma}_{crS} = \tilde{\sigma}_{crS}{}^{NC} + \tilde{\sigma}_{crS}{}^{FC}$<br><br>Cell shear stress caused by natural convection $\tilde{\sigma}_{crS}{}^{NC}$:<br>$\vec{n} \cdot \tilde{\sigma}_{crS}{}^{NC} \cdot \vec{t} = \vec{\nabla} \gamma_{cM} \cdot \vec{t} + \vec{\nabla} \gamma_c \cdot \vec{t}$<br><br>Shear stress caused by forced convection $\tilde{\sigma}_{crS}{}^{FC}$:<br>$\tilde{\sigma}_{crS}{}^{FC} = \tilde{\sigma}_{crS}{}^{CCM} + \tilde{\sigma}_{cS}{}^G$ | $\tilde{\sigma}_{MrS} = \tilde{\sigma}_{MrS}{}^{FC}$<br><br>Matrix shear stress caused by forced convection $\tilde{\sigma}_{MrS}{}^{FC}$:<br>$\tilde{\sigma}_{MrS} = \tilde{\sigma}_{MrS}{}^{TR} + \tilde{\sigma}_{MS}{}^G$ |

where $\tilde{I}$ is the unity tensor, $\vec{t}$ is the tangent vector of the biointerface, $\vec{n}$ is the normal vector of the biointerface, $\Delta p_{c \to M}$ is the component of isotropic stress, $\tilde{\sigma}_{crV}{}^{CCM}$ and $\tilde{\sigma}_{crS}{}^{CCM}$ are the cell normal and shear residual stress caused by collective cell migration, $\tilde{\sigma}_{cV}{}^G$ and $\tilde{\sigma}_{cS}{}^G$ are the cell normal and shear residual stress generated within cell cluster as a consequence of the action of the gravitational force, $\tilde{\sigma}_{MrV}{}^{TR}$ and $\tilde{\sigma}_{MrS}{}^{TR}$ are the matrix normal and shear stress caused by cell tractions.

The cell normal residual stress consists of isotropic and deviatoric parts. The isotropic part of the stress is caused by the work of cell-matrix interfacial tension expressed by the Young-Laplace equation. The interfacial tension exerts work along the biointerface area between cell cluster and matrix in order to minimize the interface. Consequently, the interfacial tension is responsible for compression and extension of both cell cluster and surrounding matrix. If the cell cluster is extended during its movement (i.e. cell wetting), this extension results in the matrix compression (Pajic-Lijakovic and Milivojevic, 2023). Otherwise, the cell cluster compression caused by de-wetting leads to the matrix strain relaxation along the cell-matrix biointerfcial area (i.e. expansion). Compression of matrix or cell cluster is labelled by sign "+", while the extension is labelled by sign "-".

The deviatoric part of the cell normal residual stress accounts for: (1) the stress caused by collective cell migration and (2) stress caused by action of the gravitational force which should be included for the case of 3D cell clusters (Pajic-Lijakovic and Milivojevic, 2023). The gravitational force can be expressed as: $\vec{f}_g = m_a \vec{g}$ (where $m_a$ is the cluster mass and $\vec{g}$ is the gravitational acceleration). For the density of cell cluster equal to $\rho_a = 1.08 \, \frac{g}{cm^3}$ (Cavicchi et al., 2018) and the radius of cell cluster equal to $r_a = 100 \: \mu m$, the corresponding force is $\vec{f}_g = 44 \: nN$ which is the same order of magnitude as the cell traction force. If the contact area between cell cluster and substrate matrix is circular with the radius of 20 $\mu m$, the corresponding interfacial area is equal to $A_a = 1.256 x 10^3 \: \mu m^2$. When $\vec{g} = (0,0, g_z)$, the corresponding z-component of normal stress is $\sigma_{gzz} = \frac{f_{gz}}{A_a}$ which is equal to 35 $Pa$. The maximum cell normal residual stress accumulation caused by 2D collective cell migration during the cell monolayer free expansion is ~150 $Pa$ (Tambe et al., 2013).

Collective cell movement induces generation of strain (volumetric and shear) which induces generation of cell stress (normal and shear), its relaxation and cell residual stress accumulation (Pajic-Lijakovic and Milivojevic, 2019;2020). Cell strain change and residual stress accumulation occur at a time scale of hours while the stress relaxation occurs at a time scale of minutes (Marmottant et al., 2009; Pajic-Lijakovic and Milivojevic, 2019).

The cell residual stress generated by collective cell migration depends primarily on the strength of cell-cell adhesion contacts. When cells establish E-cadherin mediated adherens junctions (AJs) and migrate in the form of strongly connected cell clusters, their rheological behaviour corresponds to viscoelastic solids (Pajic-Lijakovic and Milivojevic, 2019;2020). In order to formulate the cell residual stress caused by collective cell migration, it is necessary to choose proper constitutive model. Experimental results on various epithelial-like model systems such as: (1) free expansion of cell monolayers (Serra-Picamal et al., 2012), (2) rearrangement of confluent cell monolayers (Notbohm et al., 2016), and (3) cell aggregate uni-axial compression between parallel plates (Marmottant et al., 2009) pointed to the Zener constitutive model. The main characteristics of the Zener model are: (1) stress relaxes exponentially under constant strain conditions (Marmottant et al., 2009), (2) strain relaxes exponentially under constant stress conditions (Marmottant et al., 2009), and (3) the corresponding cell residual stress is purely elastic (Serra-Picamal et al., 2012; Notbohm et al., 2016). The stress relaxation occurs at a time scale of minutes, while strain relaxation, which takes place via collective cell migration, occurs at a time scale of hours (Marmottant et al., 2009; Pajic-Lijakovic and Milivojevic, 2019). Ability of strain to relax is the one of the main characteristics of viscoelastic solids (Pajic-Lijakovic, 2021). The cell residual stress accumulation caused by free expansion of cell monolayers and the rearrangement of confluent cell monolayers correlates with the corresponding strain which pointed out to elastic nature of the cell residual stress (Serra-Picamal et al., 2012; Notbohm et al, 2016). The Zener model is expressed as:

$$\tilde{\sigma}_{ci}(\Re, t_s, \tau)^{CCM} + \tau_{Ri} \dot{\tilde{\sigma}}_{ci}(\Re, t_s, \tau) = E_{ci}\tilde{\varepsilon}_{ci}(\Re, \tau) + \eta_{ci}\dot{\tilde{\varepsilon}}_{ci}(\Re, \tau) \quad (2)$$

where subscript $i$ is shear for $i \equiv S$ and normal (volumetric) for $i \equiv V$, $\Re = \Re(x, y, z)$ is the coordinate of the biointerface, $t_s$ is the short time scale (i.e. a time scale of minutes), $\tau$ is the long time scale (i.e. a time scale of hours), $\tilde{\sigma}_{ci}(\Re, t_s, \tau)^{CCM}$ is the cell stress (shear and normal), $\tilde{\varepsilon}_{ci}(\Re, \tau)$ is cell strain (shear and volumetric) caused by collective cell migration, $\tau_{Ri}$ is the corresponding stress relaxation time, $\dot{\tilde{\sigma}}_{ci}(\Re, t_s, \tau)$ is the rate of stress change, and $\dot{\tilde{\varepsilon}}_{ci}$ is the strain rate, $E_{ci}$ is the module of elasticity, and $\eta_{ci}$ is the shear or bulk viscosity. The stress relaxation occurs via successive short time relaxation cycles under constant strain per single short time cycle, while the strain change via collective cell migration occurs at a long time scale (Pajic-Lijakovic and Milivojevic, 2019; 2020). The corresponding cell stress relaxation can be expressed starting from the initial conditions at the strain $\tilde{\varepsilon}_{c0i}(\Re, \tau)$, the initial cell stress is $\tilde{\sigma}_{ci}(\Re, t = 0, \tau) = \tilde{\sigma}_{c0i}$ as: $\tilde{\sigma}_{ci}(\Re, t = 0, \tau) = \tilde{\sigma}_{c0i}$ as: $\tilde{\sigma}_{ci}(\Re, t_s, \tau) = \tilde{\sigma}_{c0i}e^{-\frac{t_s}{\tau_{Ri}}} + \tilde{\sigma}_{cRi}(\Re, \tau)\left(1 - e^{-\frac{t_s}{\tau_{Ri}}}\right)$, while the cell residual stress is elastic and equal to $\tilde{\sigma}_{cRi}(\Re, \tau) = E_{ci}\tilde{\varepsilon}_{c0i}$.

The matrix normal residual stress consists of isotropic and deviatoric parts. The isotropic part of the stress is caused by the work of the cell-matrix interfacial tension (Pajic-Lijakovic and Milivojevic, 2023). While cell expansion leads to matrix compression, the compression of cell cluster results in the matrix extension. The deviatoric part of the matrix stress accounts for: (1) the stress caused by cell tractions and (2) stress caused by action of the gravitational force which should be included for the case of 3D cell clusters (Pajic-Lijakovic and Milivojevic, 2023).

The cell shear residual stress includes two contributions: (1) the cell stress generated by the natural convection and (2) the cell stress generated by forced convection (Pajic-Lijakovic and Milivojevic, 2022c; Pajic-Lijakovic et al., 2022e). The natural convection is caused by the gradient of cell-matrix interfacial tension $\vec{\nabla}\gamma_{cM}$ established along the interfacial area and the gradient of $\vec{\nabla}\gamma_c$ established along the cell cluster itself. The gradient of interfacial tension can be expressed as $\frac{\Delta\gamma_{cM}}{\Delta L}$ (where $\Delta\gamma_{cM}$ is the interfacial tension difference and $\Delta L$ is the distance in which this gradient exist). If we suppose that the interfacial tension difference corresponds to $\Delta\gamma_{cM} \approx 1 \frac{mN}{m}$ and the distance is $\Delta L \approx 100\ \mu m$, this gradient of the interfacial tension corresponds to a cell shear stress of $\sim 10\ Pa$. It is a large value when we keep in mind that the shear stress of several tens of Pa can induce inflammation of epithelial cells (Pitanes et al., 2018). The cell movement, which results in a generation of the cell shear stress, occurs from the regions of lower cell surface tension and cell-matrix interfacial tension to the regions of higher cell surface tension and cell-matrix interfacial tension. It is a part of the Marangoni effect (Pajic-Lijakovic and Milivojevic, 2022c). The difference in the cell surface tension between leader and follower cells represents additional physical mechanism responsible for the movement of follower cells toward the leader cells. Accordingly with the fact that the cell-matrix interfacial tension is larger at the cluster front and decrease toward the cluster rear, i.e. $\gamma_{cM}^F > \gamma_{cM}^R$, this interfacial tension gradient also stimulates cell movement from the cluster rear toward the front region. The cell shear stress caused by forced convection includes two contributions: (1) the cell shear stress caused by collective cell migration and (2) the cell shear stress caused by action of the gravitational force which should be accounted for in the case of larger 3D cell cluster (Pajic-Lijakovic and Milivojevic, 2023). The maximum cell shear stress caused by free expansion of cell monolayers is equal to $\sim 100\ Pa$ (Tambe et al., 2013).

The matrix shear stress is caused actively by cell tractions and passively by the action of the gravitational force. Both contributions are generated by forced convection.

After defining the relevant physical parameters responsible for: (1) cell self-rearrangement within the cluster, (2) cell cluster wetting and de-wetting on the substrate matrix, and (3) directional cell migration, the biophysical model in the form of cell force and mass balances can be formulated.

### 3. Biophysical model

The biophysical model for movement of cell cluster on collagen I matrix is formulated in order to describe interplay among various model parameters such as cell and matrix surface tensions, interfacial tension between them, as well as the gradients of surface and interfacial tensions and cell and matrix residual stress accumulations in the form of cell force and mass balances. These parameters represent a product of dilational and volumetric viscoelasticity of both cell cluster and substrate matrix. The main goal of this modelling consideration is to elucidate the role of these physical parameters in the complex dynamics of cell migration rather than to provide exact calculations because some model parameters such as the cell-matrix interfacial tension and the gradients of surface and interfacial tensions haven't been measured yet.

## 3.1 The force balance

The force balance includes the forces which drive cell movement and the forces which reduce it. Competition between driving and resistive forces leads to an oscillatory change of cell velocity and relevant rheological parameters such as cell strain and resulted cell residual stress accumulation (Pajic-Lijakovic and Milivojevic, 2020). This oscillatory trend of cell spreading accounts for: successive, long-time oscillatory extension and compression of multicellular system and oscillations of the velocity of the cluster centre of mass (Serra-Picamal et al., 2012; Notbohm et al., 2016; Beaune et al., 2018). The oscillatory trend of collective cell migration has been recognized experimentally in various multicellular systems such as: (1) free expansion of cell monolayers (Serra-Picamal et al., 2012), (2) rearrangement of confluent cell monolayers (Notbohm et al., 2016), (3) cell aggregate rounding after uni-axial compression between parallel plates (Pajic-Lijakovic and Milivojevic, 2022b), and (4) fusion of two cell aggregates (Pajic-Lijakovic and Milivojevic, 2022b). The phenomenon has been discussed in the context of mechanical waves (Serra-Picamal et al., 2012; Notbohm et al., 2016; Pajic-Lijakovic and Milivojevic, 2020). The underlying physical mechanisms behind various types of cell movement such as: (1) the movement of cells within the cluster, (2) cell cluster wetting and de-wetting, and (3) directional cell migration are discussed in the context of various forces which were formulated based on proposed physical parameters. The roles of various forces in this complex dynamics of cell rearrangement are presented in **Table 2**.

**Table 2**. The volumetric forces which influence: (1) cell rearrangement within cluster, (2) cell cluster wetting/de-wetting, and (3) cell movement persistence

| Volumetric force | Role in cell rearrangement and migration | ref |
|---|---|---|
| Mixing force $\vec{F}_{mix}^{c-M} = \frac{1}{h_c}\vec{\nabla}_s(\gamma_{cM} - \gamma_c)$ | The thermodynamic mixing force accounts for cumulative effects of cell-matrix interactions along the biointerface. This force influences cell cluster wetting/de-wetting. | Pajic-Lijakovic and Milivojevic, 2023) |
| Marangoni force for cells $\vec{F}_M^c = \frac{1}{h_c}\vec{\nabla}_s\gamma_c$ | This force guides the self-rearrangement of cells within the cluster from the region of lower surface tension (central part of the cluster) to the region of larger surface tension (the cluster front). | This force is formulated here. |
| Matrix surface force $\vec{F}_{STF}^M = \frac{1}{h_M}\vec{\nabla}_s\gamma_M$ | This force guides directional cell movement from the region of lower matrix surface tension to the region of larger matrix surface tension. | This force is formulated here. |
| Total viscoelastic force $\vec{F}_{Tve} = \vec{F}_{ve}^c - \vec{F}_{ve}^M$ Viscoelastic force of cells $\vec{F}_{ve}^c = \vec{\nabla}\tilde{\sigma}_{cr}$ Viscoelastic force of matrix $\vec{F}_{ve}^M = \vec{\nabla}\tilde{\sigma}_{Mr}$ | The total viscoelastic force $\vec{F}_{Tve}$ accounts for the viscoelasticity of cell cluster and viscoelasticity of matrix. While viscoelastic force of cells $\vec{F}_{ve}^c$ acts to reduce cell movement, the viscoelastic force of matrix $\vec{F}_{ve}^M$ stimulates the directional cell migration by establishment the matrix stiffness gradient (i.e. durotaxis). | Murray et al. (1988) and Pajic-Lijakovic and Milivojevic(2020) |
| Interfacial tension force $n_c\vec{F}_{it}^{c-M} = n_c S^{c-M}\vec{u}_c$ | This force induces cell cluster expansion (wetting) and compression (de-wetting) depending on the value of the spreading coefficient $S^{c-M}$. | Pajic-Lijakovic et al. (2023) |
| Traction force $\rho_{c-M}\vec{F}_{tr}^{c-M} = \rho_{c-M}k_c\vec{u}_M^c$ | This is restrictive force which depends on the density and strength of FAs. | Murray et al. (1988) |
| Gravitational force $\vec{F}_g = \rho_a\vec{g}$ | This force enhances wetting of the cell cluster for the case of larger 3D cluster. | Pajic-Lijakovic and Milivojevic, 2023 |

where $h_c$ is the cell cluster height, $h_M$ is the thickness of the collagen layer, $n_c$ is the cell packing density, and $\vec{u}_M^c$ is the matrix displacement field caused by cell traction, $\rho_{c-M}$ is the density of FAs, $k_c$ is the elastic constant per single FA, $\vec{u}_c$ displacement field of cells within cluster, $\tilde{\sigma}_{cr}$ is the cell residual stress which accounts for normal and shear contributions expressed in Table 1, $\tilde{\sigma}_{Mr}$ is the matrix residual stress which accounts for normal and shear contributions expressed in Table 1, $\vec{\nabla}_s$ is the surface gradient, and $\vec{\nabla}$ is the volumetric gradient, $\rho_a$ is the cluster density, and $\vec{g}$ is the gravitational acceleration.

The mixing force $\vec{F}_{mix}^{c-M}$ represents the product of thermodynamics energetic effect of mixing of two soft matter systems such as: cell cluster and collagen I matrix. This energetic effect is caused by cell-matrix interactions along the biointerface which have a feedback on cell-cell cohesion. This force drives cell cluster: (1) wetting for $\gamma_c < \gamma_{cM}$ and (2) de-wetting for $\gamma_c > \gamma_{cM}$ and also reduces cell self-rearrangement within the cluster (Pajic-Lijakovic and Milivojevic 2023).

The Marangoni force for cells $\vec{F}_M^c$ drives cell rearrangement within the cell cluster from the region of lower cell surface tension within the cluster interior to the region of larger cell surface tension within the cluster front region. The phenomenon represents a part of the Marangoni effects (Pajic-Lijakovic and Milivojevic, 2022c).

Interfacial tension force $n_c\vec{F}_{it}^{c-M}$ (where $n_c$ is the cell packing density) drives cell wetting or de-wetting depending on the inter-relation between cell and matrix surface tensions accompanied by the interfacial tension between them expressed in the form of the cell spreading factor (Pajic-Lijakovic et al., 2022e).

While the Marangoni force of cells $\vec{F}_M^c$ and interfacial tension force $n_c\vec{F}_{it}^{c-M}$ stimulate cell movement, the viscoelastic force of cells $\vec{F}_{ve}^c$ acts to reduce cell movement through the cell residual stress accumulation (Pajic-Lijakovic and Milivojevic, 2020). Induced decrease in cell velocity results in a decrease in the cell residual stress which stimulates cell movement again. This force, which depends on the viscoelasticity of cell cluster, is responsible for oscillatory wetting and de-wetting.

Besides the interfacial tension force, the gravitational force $\vec{F}_g$, if exists, drives wetting of cell cluster. This force can be significant in the case of larger, spherical clusters (Pajic-Lijakovic and Milivojevic, 2023).

The surface and volumetric structural changes caused by cell movement induces generation of the matrix surface force $\vec{F}_{STF}^M$ and the viscoelastic force of matrix $\vec{F}_{ve}^M$, respectively which are responsible for the directional cell movement. While the matrix surface force is related to the matrix surface tension gradient, the viscoelastic force for matrix is related to the gradient of the matrix residual stress and on that base to the establishment of matrix stiffness gradient. The force $\vec{F}_{STF}^M$ causes movement of disconnected collagen fibers from the region of the lower matrix surface tension to the region of larger matrix surface tension by natural convection. This movement of the fibers has a feedback on cell cluster movement itself. Both forces should be included in the directional movement of cell clusters on collagen I gel (Clark et al., 2022). In the case of some other protein matrix, such as fibrinogen matrix, directional cell movement can be induced only by the matrix stiffness gradient, while the gradient of the matrix surface tension cannot be established by cell tractions. It is in accordance with fact that the fibrinogen surface tension is not sensitive to changes in the protein concentration (Gudapati et al., 2020).

Traction force $\rho_{c-M}\vec{F}_{tr}^{c-M}$ restricts cell movement. The phenomenon is pronounced for higher density of FAs and higher strength of single FAs (Fuhrmann et al., 2017).

The force balance for movement of cell cluster on collagen I gel can be expressed by modified model proposed by Pajic-Lijakovic and Milivojevic (2023) as:

$$\langle m \rangle_c n_c(\Re,\tau)\frac{D\vec{v}_c(\Re,\tau)}{D\tau} = \vec{F}_{mix}^c + \vec{F}_M^{c-M} + \vec{F}_{STF}^M + n_c\vec{F}_{it}^{c-M} + \vec{F}_g - \vec{F}_{Tve}^{c-M} - \rho_{c-M}\vec{F}_{tr}^{c-M} \quad (3)$$

where $\Re = \Re(x,y,z)$ represents the coordinate of cells within the aggregate-substrate contact area, $n_c$ is the packing density of cells, $\vec{v}_c(\Re,\tau)$ is cell velocity equal to $\vec{v}_c = \vec{v}_c^{CM} + \vec{v}_c{'}$, $\vec{v}_c^{CM}$ is the velocity of the cluster centre of mass, $\vec{v}_c{'}$ is the velocity of cells within the cluster which depends of cell movement within the cluster and cell cluster wetting/de-wetting, $\langle m \rangle_c$ is the average mass of single cell, $\tau$ is the time scale of hours, and $\frac{D\vec{v}_c}{D\tau} = \frac{\partial \vec{v}_c}{\partial \tau} + (\vec{v}_c \cdot \vec{\nabla})\vec{v}_c$ is the material derivative (Bird et al., 1960).

## 3.2 The mass balance

It is necessary to formulate the mass balance for cells and the mass balance for collagen matrix under the cell cluster in the form of the system of modeling equations. The mass balance of cells accounts for several contributions such as: convective or conductive flux, the Marangoni fluxes, and various taxis fluxes responsible for the directional movement of cell cluster. When cell velocity $\vec{v}_c$ is higher than diffusion cell velocity expressed as: $\vec{v}_c^d = D_{eff} \frac{1}{L_{cmax}}$ (where $D_{eff}$ is the effective cell diffusion coefficient and $L_{cmax}$ is the cell persistence length), cell migration occurs via convective mechanism (Pajic-Lijakovic and Milivojevic, 2021). While the cell rearrangement near jamming corresponds to a conductive mechanism (Nnetu et al., 2013), the convective mechanism corresponds to cell movement during: (1) free expansion of cell monolayers (Nnetu et al., 2012; Serra-Picamal et al., 2012) and (2) rearrangement of confluent cell monolayers (Notbohm et al., 2016). The effective diffusion coefficient are in the range of $\sim 0.40 \frac{\mu m^2}{min}$ - $0.10 \frac{\mu m^2}{min}$ for the cell packing density of MDCK cells in the range of $\sim 1.40 x 10^5 \frac{cells}{cm^2}$ to $\sim 2.63 x 10^5 \frac{cells}{cm^2}$, respectively (Angelini et al., 2011). The maximum correlation length for 2D collective cell migration is $L_{max} \sim 150\ \mu m$ (Petrolli et al., 2021). Corresponding diffusion cell velocity is in the range of $\vec{v}_c^{cd} = 0.005 - 0.010\ \frac{\mu m}{min}$ (Pajic-Lijakovic and Milivojevic, 2021). This diffusion velocity has been measured within cellular systems near jamming (Nnetu et al., 2013). Consequently, movement of cell cluster on collagen I matrix satisfies the condition that $\vec{v}_c > \vec{v}_c^d$ and can be described by the convective flux expressed as: $\vec{J}_{conv} = n_c \vec{v}_c$ (Murray et al., 1988).

While the convective cell flux describes movement of whole cluster, the Marangoni flux induces two contributions: (1) the cell movement within the cluster guided by the cell surface tension gradient $\vec{\nabla}_s \gamma_c$ which is formulated here and (2) cell aggregate wetting/de-wetting guided by the difference between cell surface tension and cell-matrix interfacial tension expressed as $\vec{\nabla}_s(\gamma_c - \gamma_{cM})$ which was formulated by Pajic-Lijakovic and Milivojevic 2023). The cell surface tension gradient guides cell movement within the cluster from the region of lower cell surface tension established within the cluster interior to the region of larger cell surface tension established within the cluster front. The gradient $\vec{\nabla}_s(\gamma_{cM} - \gamma_c)$ guides cell cluster: (1) wetting for the condition that $\gamma_c < \gamma_{cM}$ and (2) de-wetting for the condition that $\gamma_c > \gamma_{cM}$ (Pajic-Lijakovic and Milivojevic 2023). The resulted Marangoni flux is formulated here as: : $\vec{J}_M = k_{M1} n_c (\vec{\nabla}_s \gamma_c) + k_{M2} n_c \vec{\nabla}_s(\gamma_c - \gamma_{cM})$ (where $k_{M1}$ is the model parameter which quantifies the mobility of cells within the cluster caused by the gradient $\vec{\nabla}_s \gamma_c$ and $k_{M2}$ is the model parameter which quantifies the mobility of cells during wetting/de-wetting caused by the gradient $\vec{\nabla}_s(\gamma_c - \gamma_{cM})$).

In the case that only surface and volumetric structural changes of matrix guide directional cell migration, without chemical and electrical stimuli, two fluxes should be included. One of them is the durotaxis flux which describe directional cell movement by following the matrix stiffness gradient.

Accordingly, with fact that the matrix stiffness gradient correlates with the gradient of the matrix residual stress, the durotaxis flux is expressed as: $\vec{J}_d = k_d n_c \Delta V_m |\vec{\nabla}\tilde{\sigma}_{Mr}|$ (where $k_d$ is the model parameter which represents a measure of the mobility of collagen fibers within the gel induced by cell tractions and $\Delta V_m$ is the volume of a matrix part) (Pajic-Lijakovic and Milivojevic, 2023). The stiffness gradient of the matrix can be induced by cells themselves during their movement.

The other flux (so called "surfo-taxis" flux), formulated here, describes directional cell movement by following the gradient of matrix surface tension and can be expressed as: $\vec{J}_{ST} = k_{ST} n_c (\vec{\nabla}_s \gamma_M(C_{col}))$ (where $k_{ST}$ is the model parameter which represents a measure of collagen-cell surface interactions which influence cell movement and $C_{col}$ is the concentration of the collagen fibers under the cell cluster).

The corresponding mass balance can be expressed by modified model proposed by Pajic-Lijakovic and Milivojevic (2023) as:

$$\frac{\partial n_c(\Re,\tau)}{\partial \tau} = -\vec{\nabla} \cdot (\vec{J}_{conv} + \vec{J}_{ST} + \vec{J}_d + \vec{J}_M) \qquad (4)$$

where $\vec{J}_{conv}$ is the convective flux, $\vec{J}_{ST}$ is the surfo-taxis flux, $\vec{J}_d$ is the durotaxis flux, and $\vec{J}_M$ is the Marangoni flux. When the directional cell movement is considered based on the proposed fluxes $\vec{J}_{ST}$ and $\vec{J}_d$, four cases can be distinguished such as: (1) $\vec{J}_{ST} \ll \vec{J}_d$, (2) $\vec{J}_{ST} \gg \vec{J}_d$, (3) $\vec{J}_{ST} \sim \vec{J}_d$ and both fluxes are oriented in the same direction, and (4) $\vec{J}_{ST} \sim \vec{J}_d$ and fluxes are oriented in the opposite directions. Which case exists in some experiment depends on cell type and physical properties of matrix such as the filament rigidity and the strange of inter- and intra- filaments interactions. The forth case could be related with negative durotaxis. However, additional experiments are needed in order to confirm this claim.

The mass balance of the disconnected collagen fibers under the cell cluster could be expressed in the form of Marangoni flux as:

$$\frac{\partial C_{col}(\Re,\tau)}{\partial \tau} = -\vec{\nabla} \cdot \vec{J}_{col\,M} \qquad (5)$$

where $\vec{J}_{col\,M}$ is the Marangoni flux for collagen fibers equal to: $\vec{J}_{col\,M} = k_{col} n_c (\vec{\nabla}_s \gamma_M(C_{col}))$ and $k_{col}$ is the the model parameter which represents a measure of collagen-cell surface interactions capable of inducing local disintegration of the matrix and the rearrangement of collagen fibers. Model eqs. 3-5 should be solved simultaneously. The collagen concentration undergo long-time change and could reach out the equilibrium value when $\vec{\nabla}_s \gamma_M \to 0$ which is possible only for higher mobility of collagen fibers. Clark et al. (2022) pointed to the chain mobility as the key factor for the establishment of movement persistence of cell cluster.

4. Conclusion

Over the last years it has become clear that movement of cell collectives along substrate matrix during morphogenesis, tissue regeneration and cancer invasion involve several inter-connected migration modes such as: cell movement within the migrating clusters, the cluster wetting/de-wetting, and directional cell movement. The matrix stiffness gradient has been recognized as a main physical property of the matrix gel responsible for cell wetting/de-wetting and directional cell movement. Cell movement within the clusters has been discussed in the context of biochemical mechanism related to cell signalling which is established between leader and follower cells, while physical mechanism which also could guide this cell movement hasn't been identified yet. The main goal of this theoretical consideration is to point out the roles of various physical parameters such as: (1) the cell and matrix surface tensions, (2) interfacial tensions between them, (3) gradients of surface and interfacial tensions, and (4) accumulation of the cell and matrix residual stresses in the establishment of various migration modes. These parameters represent a product of dilational and volumetric viscoelasticity of both, cell cluster and the hydrogel matrix which are simultaneously changing caused by cell-matrix interactions along the biointerface. Cumulative effects of these interactions stimulate internal cellular mechanisms such as cell signalling and gene expression and also influence the physical parameters which have a feedback on single cell state. This complex cause-consequence relationship is discussed based on the biophysical model which includes cell force and mass balances formulated at a supracellular level. Following conclusions can be extracted:

- The cell surface tension gradient established between leader and follower cells guides cell movement from the region of lower cell surface tension (characteristic for the follower cells) to the region of larger cell surface tension (characteristic for the leader cells). This phenomenon represents a part of the Marangoni effect which also exists in various soft matter systems.
- Cell cluster wetting/de-wetting depends on interplay between physical parameters such as cell and matrix surface tensions accompanied by the interfacial tension between them through the cell spreading factor. The gradient of the difference between cell surface tension and cell—matrix interfacial tension is included in the mixing force (the cell force balance) and in the Marangoni flux (the cell mass balance).
- Surface and volumetric structural changes of matrix, caused by movement of cell collectives, are responsible for the directional cell migration. The volumetric structural changes lead to the inhomogeneous accumulation of the matrix residual stress and on that base result in the establishment of the matrix stiffness gradient, while the surface structural changes can induce the generation of the matrix surface tension gradient. Both gradients can appear as a product of cell movement depending on the physical properties of the matrix such as: the rigidity of filaments and the strange of inter- and intra-filaments interactions.

Additional experiments are necessary to correlate: (1) the matrix surface tension gradient with the cell persistence length, (2) cell-matrix interfacial tension with the cell surface tension, (3) cell-matrix interfacial tension with the cell normal residual stress in one hand and with the matrix residual stress in the other.

**Author contributions**: All authors contributed equally to the paper.

**Acknowledgment**: IPL and MM are supported by the Ministry of Education, Science and Technological Development of the Republic of Serbia (Contract No. 451-03-68/2022-14/200135). We thank Andrew G. Clark (University of Stuttgart) for discussions which inspired this work and useful comments which help us to finalize the manuscript.

**Declaration of interest**: The authors report no conflict of interest.